\documentclass[prl, aps, twocolumn, superscriptaddress, nofootinbib, tightenlines, nobibnotes, showpacs]{revtex4}

\usepackage{bigints}
\usepackage{amssymb}
\usepackage{graphicx}
\usepackage{dcolumn}
\usepackage{bm}
\usepackage{multirow}
\usepackage{hyperref}
\usepackage{comment}
\usepackage{xcolor}
\usepackage[caption=false]{subfig}

\definecolor{lapislazuli}{rgb}{0.15, 0.38, 0.61}
\definecolor{YKblue}{rgb}{0.0, 0.18, 0.65}
\definecolor{carmine}{rgb}{0.81, 0.09, 0.13}
\definecolor{lavender}{rgb}{0.84, 0.79, 0.87}

\hypersetup{
colorlinks=true,
linkcolor=carmine,
linktoc=page,
citecolor=lapislazuli,
urlcolor=lapislazuli
}

\everymath{\displaystyle}

\begin{document}

\preprint{APS/123-QED}

\title{Terahertz Laser Combs in Graphene Field-Effect Transistors}

\author{Pedro Cosme}
\email{pedro.cosme.e.silva@tecnico.ulisboa.pt} 
\affiliation{Instituto de Plasmas e Fus\~ao Nuclear, Lisboa, Portugal}
\affiliation{Instituto Superior T\'ecnico, Lisboa, Portugal}

\author{Hugo Ter\c{c}as}
\email{hugo.tercas@tecnico.ulisboa.pt} 
\affiliation{Instituto de Plasmas e Fus\~ao Nuclear, Lisboa, Portugal}
\affiliation{Instituto Superior T\'ecnico, Lisboa, Portugal}

\begin{abstract}
Electrically injected terahertz (THz) radiation sources are extremely appealing given their versatility and miniaturization potential, opening the venue for integrated-circuit THz technology. In this work, we show that coherent THz frequency combs in the range $0.5~\mathrm{THz}<\omega/2\pi<10~\mathrm{THz}$ can be generated making use of graphene plasmonics. Our setup consists of a graphene field-effect transistor with asymmetric boundary conditions, with the radiation originating from a plasmonic instability that can be controlled by direct current injection. We put forward a combined analytical and numerical analysis of the graphene plasma hydrodynamics, showing that the instability can be experimentally controlled by the applied gate voltage and the injected current. Our calculations indicate that the emitted THz comb exhibits appreciable temporal coherence ($g^{(1)}(\tau)>0.6$) and radiant emittance ($10^{7}\,\mathrm{Wm^{-2}}$). This makes our scheme an appealing candidate for a graphene-base THz laser source. Moreover, a mechanism for the instability amplification is advanced for the case of substrates with varying electric permitivitty, which allows to overcome eventual limitations associated with the experimental implementation. 

\end{abstract}

\maketitle


{\it Introduction.}$-$Terahertz (THz) radiation consists of electromagnetic (EM) waves within the frequency range from 0.1 to 10 THz, filling the gap between microwave and infrared light. The technology for its production is currently a very active field of research \cite{Williams2007,editorial2013}, since THz radiation has numerous applications, comprising sensing, imaging, metrology and spectroscopy \cite{Woolard2003a,Pornsuwancharoen2013}. A major reason for the hype around THz radiation is that fact that it is able to penetrate several materials that are opaque to visible and IR radiation, while the short wavelength provides high image resolution \citep{stantchev2017}. On the other hand the attenuation in water can provide information for medical imaging while being non-ionising and biologically safe.  \par

Among all forms of THz radiation, THz laser (THL) combs play a prominent role within such technology \cite{Fuser2014,Barmes2013}. However, THL generation still faces significant difficulties, being restricted to gas lasers \cite{Dai2011,Dai2009}, with low efficiency, quantum cascade lasers \cite{Burghoff2014,Williams2007}, which require extremely low temperatures, and free electron lasers \cite{Tan2012}, practically impossible to miniaturize. With the advent of graphene plasmonics, new techniques relying in optical pumping have been put forward \cite{AltaresMenendez2017a,Outsuji2013,Otsuji2014b}. The progress in graphene based transistors \cite{Schwierz2010} paved the way to the possibility for all-electrical miniaturized devices for low power radiation emission and detection. Yet, such integrated-circuit THz technology based on graphene is at its infancy, notwithstanding some experimental studies in visible and mid-infrared light emission \cite{Kim2015,Kim2018,Beltaos2017}. More recently, THz emission from dual gate graphene field-effect transistor (FET) due to electron/hole recombination in a p-i-n junction has been made possible \cite{Yadav2017a,Yadav2018a}. Practical solutions towards inexpensive, compact and easy-to-operate lasing devices are, therefore, desirable. \par

In this Letter, we exploit a scheme for the generation of coherent THz frequency combs in graphene field-effect transistors, arising from the Dyakonov-Shur (DS) plasmonic instability \cite{Dyakonov1993,Kargar2018}. The latter can be excited via the injection of an electric current, thus forgoing the necessity of optical pumping. This opens the possibility to the development of an all-electric, low-consumption stimulated THL, capable of operating at room temperature. Our calculations are based on the hydrodynamic formulation of the plasma in monolayer graphene, from which we analytically obtain the instability criteria and numerically extract the radiation spectrum, intensity and correlation. Our findings reveal that the emitted THL comb exhibits appreciably large values of both spatial and temporal coherences, suggesting our scheme to be a competitive solution towards THz laser light with integrated-circuit technology. Finally, a new passive mechanism for the amplification of the DS instability in gated graphene is introduced.

\begin{figure}[!t]
\subfloat[\label{sfig:FET}]{
\includegraphics[]{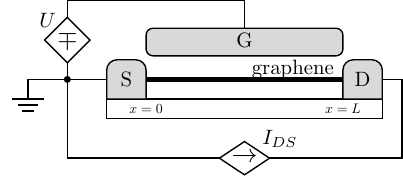}
}\hfill
\subfloat[\label{sfig:n_ex}]{
\includegraphics[]{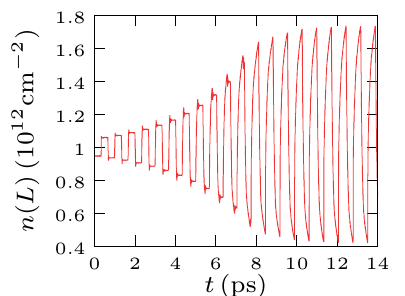}
}
\caption{Left panel: Schematic diagram for gated graphene transistor. For the DS instability to occur a fixed current  $I_{DS}$ is injected at the drain while mantaining the electronic density of the source constant. Right panel: Example of the instability growth of electronic density at the drain.}
\label{fig:dupla}
\end{figure}


{\it Graphene plasma instability.}$-$The dynamics of the electronic flow in a graphene FET can be described with the help of a hydrodynamic model \cite{Chaves2017, Muller2009, Svintsov2013}. Assuming the transport to be restricted to one (say $x$) direction, the latter reads
\begin{equation}
\begin{array}{c}
\frac{\partial n}{\partial t} +\frac{\partial}{\partial x}nv = 0,\\\\
\frac{\partial v}{\partial t} + v \frac{\partial v}{\partial x} = \frac{F}{m^*} -\frac{1}{m^*n}\frac{\partial P}{\partial x}, 
\end{array}
\label{eq_model}
\end{equation}
where $n$ is the 2D electronic density, $v$ is the flow velocity, $F$ is the total force exerted in the electrons, $P=\hbar v_F\sqrt{\pi n^3}/3$ is the pressure (with $v_F\sim 10^6~\mathrm{ms^{-1}}$ denoting the Fermi speed), and $m^*$ the electron effective mass. The validity of the hydrodynamic model in Eq. \eqref{eq_model} is granted thanks to the large value of the mean free path of electron scattering with phonons and impurities $l$ at room temperature ($l>0.2\,\mathrm{\mu m}$), thus assuring that ballistic transport holds. We also consider graphene to be in the degenerate Fermi liquid regime, provided that the Fermi level remains bellow the Van Hove singularities. At room temperature, the latter holds for Fermi energies in the range $0.025~\mathrm{eV}\ll E_F\ll 3~\mathrm{eV}$. 
 
The fact that electrons in graphene behave as massless fermions poses a difficulty to the development of hydrodynamic models with explicit dependency on the mass. Here, the Drude mass $m^*= \hbar\sqrt{\pi n_0}/v_F$, with $n_0$ denoting the equilibrium carrier density, is used as an effective mass \cite{Svintsov2013,Neto2007,Chaves2017}. In the field-effect transistor (FET) configuration comprising a drain, a source and a gate (see Fig. \ref{sfig:FET} for a schematic representation), the electric force exerted on the electrons is dominated by the external potential that screens Coulomb interaction. The applied bias potential $U$ has the contribution of both the parallel plate capacitance $C_g=\varepsilon/d_0$ (with $d_0$ denoting the distance between the gate and the graphene sheet) and the quantum capacitance \cite{Zhu2009,Fang2007,Droscher2012,Sarma2010}  $C_q=2e^2\sqrt{\pi n}/\pi \hbar v_F$, as 
 \begin{equation}
U=en\left(\frac{1}{C_g}+\frac{1}{C_q}\right).
 \end{equation}
For carrier densities in the range $n\gtrsim10^{12}\,\mathrm{cm}^{-2}$, quantum capacity dominates, and the potential can be approximated as $U\simeq end_0/\varepsilon$. Keeping the pressure term up to first order in the density, the fluid model in \eqref{eq_model} can be recast in a dimensionless form as 
\begin{equation}
\frac{\partial n}{\partial t} +\frac{\partial}{\partial x}\left(nv\right) = 0, \quad \frac{\partial v}{\partial t} + v \frac{\partial v}{\partial x} +    \frac{S^2}{v_0^2}\frac{\partial n}{\partial x}=0,
\label{eq:nondimHydro}     
\end{equation}
where $v_0$ is the electron mean drift velocity along the graphene channel and  $S^2\equiv e^2d_0n_0/(m^*\varepsilon )+v_F^2/2$ can be interpreted as sound velocity of the carriers fluid, as the dispersion relation for the electron fluctuations, $\omega=(v_0\pm S)k$,  is similar to that of a shallow water. For typical values, the ratio $S/v_0$ scales up to a few tens. 


The hydrodynamic model in Eq. \eqref{eq:nondimHydro} contains an instability under the boundary conditions of fixed density at source $n(x=0)=n_0$ and fixed current density at the drain $n(x=L)v(x=L)=n_0v_0$, dubbed in the literature as the Dyakonov-Shur (DS) instability \cite{Dyakonov1993,Dyakonov2011}. The later arises from the multiple reflections of the plasma waves at the boundaries, which provides a positive feedback for the incoming waves driven by the current at the drain. Combining Eq. \eqref{eq:nondimHydro} with the asymmetric boundary conditions described above, the dispersion relation becomes complex, $\omega=\omega_r+i\gamma$, where $\omega_r$ is the electron oscillation frequency and $\gamma$ is the instability growth rate \cite{Dyakonov1993,Dmitriev1997,Crowne1997} 
\begin{equation}
\begin{array}{c}
    \omega_r = \frac{|S^2-v_0^2|}{2LS}\pi, \\\\
    \gamma = \frac{S^2-v_0^2}{2LS}\log\left|\frac{S+v_0}{S-v_0}\right|.
\end{array}
\label{eq:omegatotal}
\end{equation}
Therefore, given the dependence of $S$ with gate voltage, and as $v_0n_0=I_{\rm DS}/We$, with $I_{\rm DS}$ representing the source-to-drain current and $W$ the transverse width of the sheet, the frequency can be tuned by the gate voltage and injected drain current, not being solely restricted to the geometric factors of the FET. 

\begin{figure}[!t]
    \centering
    \includegraphics[]{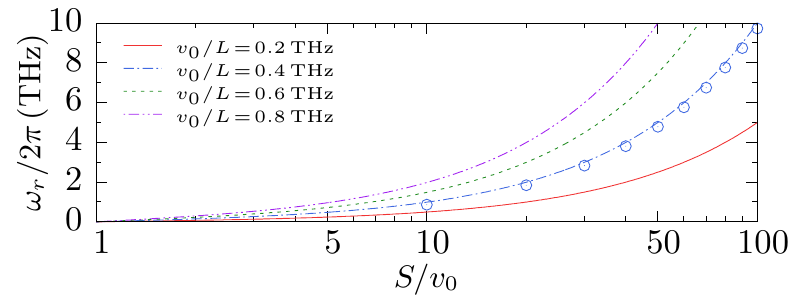}
    \includegraphics[]{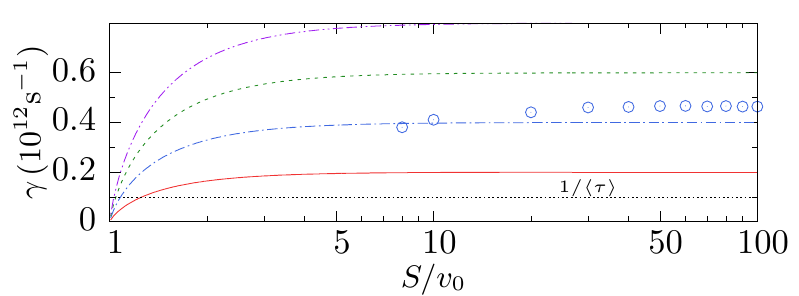}
    \caption{Top panel: Frequency of first mode vs. $S/v_ 0$, the parameter controlling the DS instability in graphene. Bottom panel: Instability growth rate vs. $S/v_ 0$. The solid and dashed lines depict the theoretical curves from \eqref{eq:omegatotal}, while the open dots the simulation results for $v_0/L=0.4~\mathrm{THz}$. The horizontal solid line line indicates the instability threshold $\gamma \tau=1$, as determined by typical experimental conditions.}
    \label{fig:Re(w)andIm(w)}
\end{figure}
After an initial transient time, the DS instability saturates due to the nonlinearities and the system goes in a cycle of shock and rarefaction waves that sustain the collective motion of the electrons (see Fig. \ref{fig:dupla}b). Evidently, such collective oscillation radiates in the same main frequency $\omega_r$ that lies in the THz range for typical values of parameters, as depicted in Fig. \ref{fig:Re(w)andIm(w)}. Moreover, as the electrons are transported along density shock waves, they bunch together and behave as a macroscopic dipole. As a consequence, the emitted radiation is highly coherent, a fact that we will demonstrate below and that we believe to be at the basis of a THL source. \par 

In experimental conditions, the ideal situation described above must be analysed with care. Indeed, for the DS instability to take place, the growth rate $\gamma$ has to be larger than the total relaxation rate, $1/\tau$, due to scattering with impurities and phonons. Recent experiments performed at room temperature point to an electron mobility of monolayer suspended graphene of $\mu\simeq 10^5\,\mathrm{cm^2V^{-1}s^{-1}}$ \cite{Bolotin2008,Zhu2009}, that results in an average relaxation rate $1/\tau \approx10^{11}\,\mathrm{s^{-1}}$. Fortunately, for a suitable choice of parameters, we can safely operate in the regime $\gamma\tau>1$, as shown in Fig. \ref{fig:Re(w)andIm(w)}.

\begin{figure}[!t]
    \centering
    \includegraphics[]{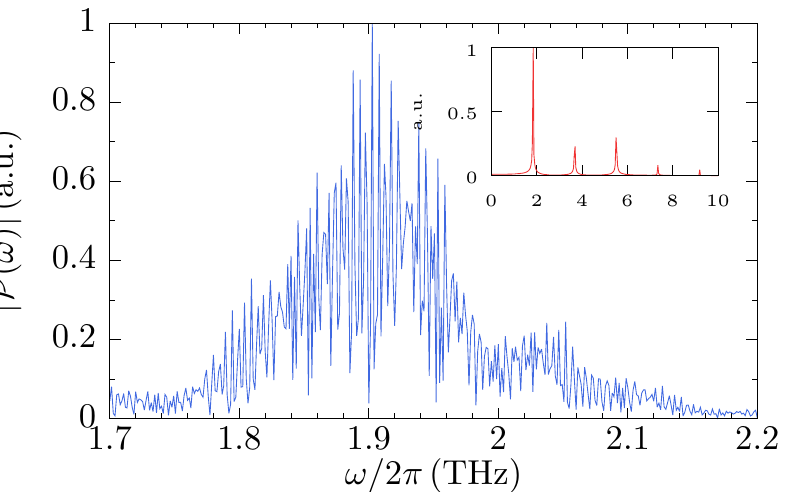}
    \caption{Frequency comb obtained for $L=0.75\,\mathrm{\mu m}$, $v_0=0.3v_F$ and $S/v_0=20$. A pulse repetition rate of $5\,\mathrm{GHz}$ with a 10\% duty cycle has been used. In the inset, the spectrum for the dc case.  Both Fourier spectra have been calculated with the FFTW3 library \cite{Frigo2005}.}
    \label{fig:FreqCombdB}
\end{figure}

{\it Numerical simulation of the THz frequency comb.}$-$The hyperbolic set of fluid equations in \eqref{eq:nondimHydro} have been integrated using a second-order (time and space) Richtmyer two-step Lax-Wendroff scheme \cite{LeVeque1992}. The suppression of numerical oscillations at the shock front has been implemented by means of a moving average filter on the spatial domain, from which the $n(x,t)$ and $v(x,t)$ profiles are obtained as well as the integrated current and tension drop across the FET. From the output current and density, the electromagnetic field can then be calculated from Jefimenko's integral equations  \cite{Griffiths1991}
\begin{equation}
\begin{array}{c}
     \mathbf{E}(\mathbf{r}, t) = \frac{en_0}{4 \pi \varepsilon_0}\!\bigintsss\!\!\!\mathrm{d}^2\mathbf{r'} \left[\frac{n}{|\mathbf{R}|^3} + \frac{\partial_t n}{|\mathbf{R}|^2 c}\right]\mathbf{R} - \frac{\partial_t (n\mathbf{v})}{|\mathbf{R}| c^2}, \\\\
     
    \mathbf{B}(\mathbf{r}, t) = \frac{e\mu_0n_0v_0}{4 \pi}\!\bigintsss\!\!\!\mathrm{d}^2\mathbf{r'} \left[\frac{n\mathbf{v}}{|\mathbf{R}|^3} + \frac{\partial_t (n\mathbf{v})}{|\mathbf{R}|^2 c} \right]\times\mathbf{R}.
\end{array}
\label{eq:jefimenko}
\end{equation}
with $\mathbf{R}=\mathbf{r}-\mathbf{r'}$ being the displacement vector and $c$ the speed of light. The direct integration of Eq. \eqref{eq:jefimenko} allows the reconstruction of the emitted fields both in near field and far field regimes.
\begin{figure}[!t]
    \centering
    \includegraphics[]{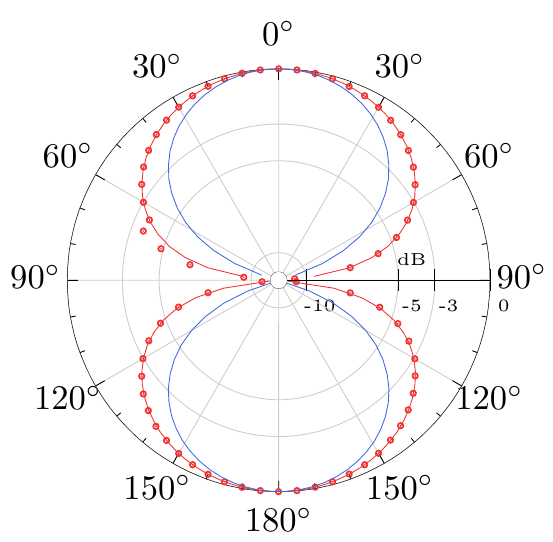}
    \caption{Far-field radiation pattern of the graphene FET emitter at a distance $R=1000L$. Average Poynting vector normalized to its maximum value vs. polar angle $\theta$ in the vertical $xz$-plane. Points from numeric simulation follow $\langle S(\theta)\rangle\propto |\cos\theta|$ closely (red solid line), whereas the radiation from a dipole antenna is narrower with $\langle S(\theta)\rangle\propto\cos^2\theta$ (blue solid line). }
    \label{fig:pattern}
\end{figure}
For the latter, our calculations provide a radiant emittance of the order of $10^{7}\,\mathrm{Wm^{-2}}$. The reconstructed radiation pattern of the graphene layer (i.e. without the reckoning radiation reflection/absorption effects due to the metallic gate, an approximation that holds as the ratio $c/v_0\sim1000$, implying the radiation wavelength $\lambda \simeq 4Lc/S$ to be much larger that the typical FET dimensions), shows a wide omnidirectional profile with a half-power beam width of $120^{\circ}$, as patent in Fig. \ref{fig:pattern}. In fact, the angular Poynting vector profile is $\langle S\rangle\propto\vert \cos\theta\vert$, unlike the typical dipolar emitter $\langle S\rangle_{\rm dip}\propto\cos^2\theta$. Such a wide lobe profile can be explained by the fact that the collective movement of charges occurs on the entire area of the graphene FET, similarly to patch antennae \cite{Kobayashi2016}. The radiated spectrum consists of a frequency comb in the THz range, with frequencies $\omega = (j+1) \omega_r$ with $j\in\mathbb{N}$, as can be seen in the inset of Fig. \ref{fig:FreqCombdB}. Hereafter, a suitable design of a resonance cavity would allow the selection and amplification of the desired mode. Furthermore, if a periodic inversion of polarity is imposed across the channel, in such a way that after the saturation time the perturbation due to DS instability decays, a train of short pulses can be formed. This can be implemented with a pulse generator controlling the injected current at drain with a repetition rate $f_{\rm rep}$. In Fourier space, such pulses form a THz frequency comb around the main frequency with frequencies given by $\omega=\omega_r+2\pi j f_{\rm rep}$ with $j\in\mathbb{Z}$ as seen in Fig. \ref{fig:FreqCombdB}. \par
\begin{figure}[!t]
    \centering
    \includegraphics[]{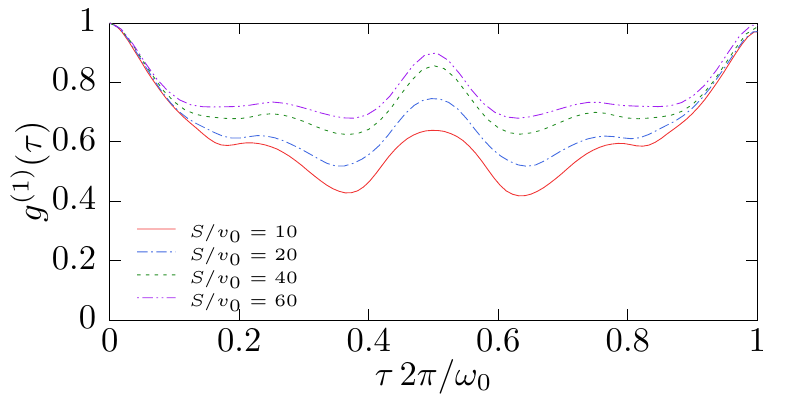}
    \includegraphics[]{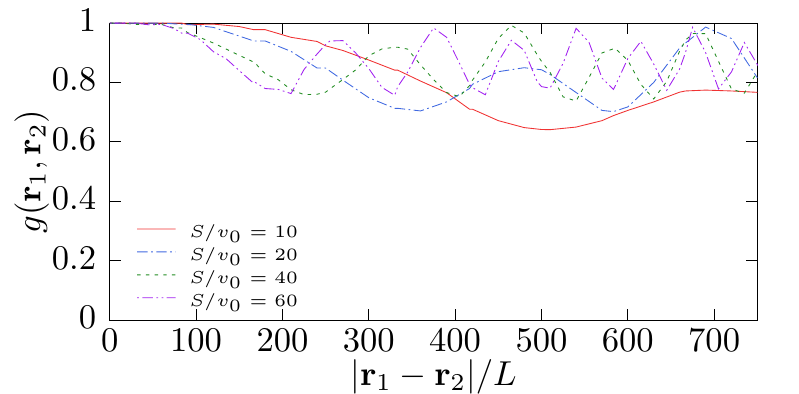}
    \caption{Top panel: Temporal first order degree of coherence vs. delay normalized to the period. Bottom panel: Spatial degree of coherence
    vs. displacement transverse to main lobe propagation.
    }
    \label{fig:coherencedegree}
\end{figure}

{\it THz laser coherences.}$-$In order to determine wether or not the emitted radiation can be understood as a THz laser, we proceed to the calculation of both the temporal and spatial coherences \cite{BornWolf1980}, 
\begin{equation}
\begin{array}{c}
    g^{(1)}(\tau) = \frac{\langle E^*(\mathbf{r},t)E(\mathbf{r},t+\tau) \rangle}{\langle E^2(\mathbf{r},t)\rangle}\quad \mathrm, \\\\
g(\mathbf{r}_1,\mathbf{r}_2) = \frac{\langle E^*(\mathbf{r}_1,t)E(\mathbf{r}_2,t) \rangle}{\sqrt{\langle E^2(\mathbf{r_1},t)\rangle\langle E^2(\mathbf{r_2},t)\rangle}}.
\end{array}
\label{eq_coherence}
\end{equation}
As expected, the radiated field exhibits appreciably large coherences, both in far field and near field (not shown) regimes, that increase with $S/v_0$, as presented in Fig. \ref{fig:coherencedegree}. As mentioned above, such an appreciable coherence degree is attributed to the collective motion of the electrons bunching together in the shock and rarefaction waves cycle. The observed oscillation in the temporal coherence is due to the frequency mixing in the frequency comb, a feature that could be suppressed with the help of a THz cavity, allowing for mode selection in the THz comb. \par
\begin{figure}
    \centering
    \includegraphics[]{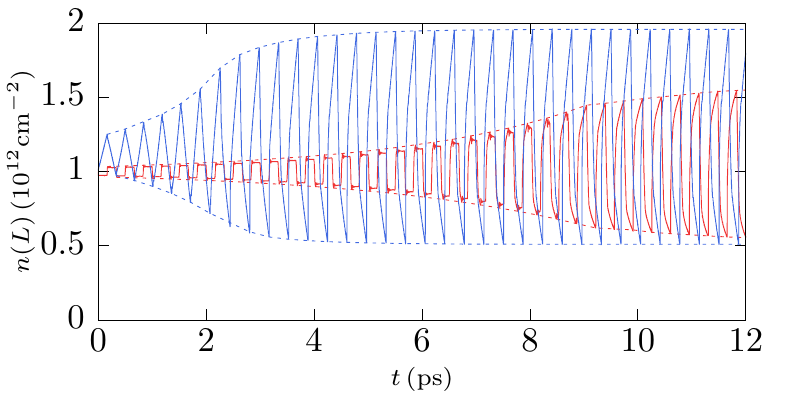}
    \caption{Temporal evolution of electronic density at the drain. Comparison of the growth rate in the case of constant $S=40$ (red line) vs. the presence of linear gradient, $S/v_0=40(1-0.05x/L)$ (blue line). An increase of $\sim 33\%$ in the electronic density at saturation is attained.}
    \label{fig:growthratecompare}
\end{figure}

{\it Improving the DS instability.}$-$As mentioned above, one possible limitation of our process may arise when we approach the threshold region $\gamma\tau=1$. To circumvent this issue without changing the setup, we found that by varying the speed of the the plasma waves along the channel length, i.e. by taking $S=S_0(1-\alpha x)$ (either by manipulating the permittivity $\varepsilon$ or the gate distance $d_0$), a remarkable enhancement of the instability growth rate can be achieved, while no significant impact on the spectrum is observable. Such modification on the velocity along the FET channel introduces a positive feedback in the current instability, which leads to a larger value of the growth rate and, consequently, to a faster saturation of the DS instability. This features are illustrated in Fig. \ref{fig:growthratecompare}. As such, the electron density at saturation is higher, resulting in a $\sim 33 \%$ increase in the emitted power for $\alpha=0.05/L$. The numerically extracted valued of the growth rate in the speed-gradient scheme, $\tilde{\gamma}$, are significantly larger than $\gamma$ in Eq. \eqref{eq:omegatotal}, as it can be stated in Table \ref{tab:sdeclive}. This mechanism can be seen as analogous to wave shoaling effect on shallow waters systems and the shock wave amplitude is likewise amplified in the presence of the velocity gradient \cite{gourlay2011}. Although out of the scope of the present work, the investigation of the DS mechanism in combination with other positive feedback configurations, such as subtract patterning \cite{zolotovskii2018} and counter flows \cite{morgado2017}, will certainly deserve our attention in the near future. 
\begin{table}[!b]
\caption{Normalized growth rate $\tilde{\gamma}/\gamma$ in the presence of a negative gradient of the local sound speed in the FET channel, $S=S_0(1-\alpha x)$.\label{tab:sdeclive}}
\begin{ruledtabular}    
    \begin{tabular}{l D{-}{\pm}{-1}  D{-}{\pm}{-1}  D{-}{\pm}{-1}  D{-}{\pm}{-1} }
        &\multicolumn{4}{c}{$S_0/v_0$}\\ 
  \multicolumn{1}{c}{$\alpha/L$}  & \multicolumn{1}{c}{20} & \multicolumn{1}{c}{40} & \multicolumn{1}{c}{60} & \multicolumn{1}{c}{80} \\ \colrule
    0.025 & 1.9-0.5 & 3.8-0.3  & 6.9-0.2 & 10.9-0.4 \\ 
      0.05 & 4.5-0.4 & 12.5-0.5 & 26-1    & 42.3-0.9 \\ 
    0.075 & 7.3-0.4 & 21.1-0.3 & 37-2    & 68-2 \\ 
    0.1  & 8.5-0.3 & 26-2     & 53-1    & 93-3 
    \end{tabular}
    \end{ruledtabular}
\end{table}


{\it Conclusion.}$-$We make use of a hydrodynamic model to describe a plasmonic instability (Dyakonov-Shur instability) taking place in a graphene field-effect transistor at room temperature. Our scheme, based on the control of the electron current at the transistor drain, results in the emission of a THz frequency comb. Numerical simulations suggest that the emitted THz radiation is extremely coherent, thus be an appealing candidate for a THz laser source. Our findings point towards a method for the development of tuneable, all-electrical THz antennae, dismissing the usage of external light sources such as THz solutions based on quantum cascade lasers. This puts graphene plasmonics and, in particular, graphene field-effect transistors in the run for competitive, low-consumption THz devices based on integrated-circuit technology, allowing the fabrication of patch arrays designed to enhance the total radiated power. \par
Additional effects can be taken into account, such as the electron-phonon coupling. This can particularly important in the case of suspended graphene, as the out-of-plane vibrations (flexural phonons) play a significant role in the electron transport \cite{castro2010}. Moreover, important effect related to electron viscosity may arise in thin graphene ribbons \cite{tomadin2014}. The later, more relevant for very small devices, may hinder the plasmonic instability and, for that reason, it is desirable to combine the Dyakonov-Shur configuration with other positive-feedback schemes. Finally, the instability amplification by resonances taking place in magnetized graphene plasmas may also conduct to interesting solutions \cite{chen2012}. \par
One of the authors (H.T.) acknowledges Funda\c{c}\~{a}o da Ci\^{e}ncia e Tecnologia (FCT-Portugal) through the grant number IF/00433/2015.

\bibliography{references.bib}

\begin{thebibliography}{45}%
\makeatletter
\providecommand \@ifxundefined [1]{%
 \@ifx{#1\undefined}
}%
\providecommand \@ifnum [1]{%
 \ifnum #1\expandafter \@firstoftwo
 \else \expandafter \@secondoftwo
 \fi
}%
\providecommand \@ifx [1]{%
 \ifx #1\expandafter \@firstoftwo
 \else \expandafter \@secondoftwo
 \fi
}%
\providecommand \natexlab [1]{#1}%
\providecommand \enquote  [1]{``#1''}%
\providecommand \bibnamefont  [1]{#1}%
\providecommand \bibfnamefont [1]{#1}%
\providecommand \citenamefont [1]{#1}%
\providecommand \href@noop [0]{\@secondoftwo}%
\providecommand \href [0]{\begingroup \@sanitize@url \@href}%
\providecommand \@href[1]{\@@startlink{#1}\@@href}%
\providecommand \@@href[1]{\endgroup#1\@@endlink}%
\providecommand \@sanitize@url [0]{\catcode `\\12\catcode `\$12\catcode
  `\&12\catcode `\#12\catcode `\^12\catcode `\_12\catcode `\%12\relax}%
\providecommand \@@startlink[1]{}%
\providecommand \@@endlink[0]{}%
\providecommand \url  [0]{\begingroup\@sanitize@url \@url }%
\providecommand \@url [1]{\endgroup\@href {#1}{\urlprefix }}%
\providecommand \urlprefix  [0]{URL }%
\providecommand \Eprint [0]{\href }%
\providecommand \doibase [0]{http://dx.doi.org/}%
\providecommand \selectlanguage [0]{\@gobble}%
\providecommand \bibinfo  [0]{\@secondoftwo}%
\providecommand \bibfield  [0]{\@secondoftwo}%
\providecommand \translation [1]{[#1]}%
\providecommand \BibitemOpen [0]{}%
\providecommand \bibitemStop [0]{}%
\providecommand \bibitemNoStop [0]{.\EOS\space}%
\providecommand \EOS [0]{\spacefactor3000\relax}%
\providecommand \BibitemShut  [1]{\csname bibitem#1\endcsname}%
\let\auto@bib@innerbib\@empty
\bibitem [{\citenamefont {Williams}(2007)}]{Williams2007}%
  \BibitemOpen
  \bibfield  {author} {\bibinfo {author} {\bibfnamefont {B.~S.}\ \bibnamefont
  {Williams}},\ }\href {\doibase 10.1038/nphoton.2007.166} {\bibfield
  {journal} {\bibinfo  {journal} {Nature Photonics}\ }\textbf {\bibinfo
  {volume} {1}},\ \bibinfo {pages} {517} (\bibinfo {year} {2007})}\BibitemShut
  {NoStop}%
\bibitem [{edi(2013)}]{editorial2013}%
  \BibitemOpen
  \href {\doibase 10.1038/nphoton.2013.239} {\bibfield  {journal} {\bibinfo
  {journal} {Nature Photonics}\ }\textbf {\bibinfo {volume} {7}},\ \bibinfo
  {pages} {665} (\bibinfo {year} {2013})}\BibitemShut {NoStop}%
\bibitem [{\citenamefont {Woolard}\ \emph {et~al.}(2003)\citenamefont
  {Woolard}, \citenamefont {Loerop},\ and\ \citenamefont
  {Shur}}]{Woolard2003a}%
  \BibitemOpen
  \bibinfo {editor} {\bibfnamefont {D.~L.}\ \bibnamefont {Woolard}}, \bibinfo
  {editor} {\bibfnamefont {W.~R.}\ \bibnamefont {Loerop}}, \ and\ \bibinfo
  {editor} {\bibfnamefont {M.~S.}\ \bibnamefont {Shur}},\ eds.,\ \href@noop {}
  {\emph {\bibinfo {title} {{Terahertz sensing technology Volume 2: Emerging
  Scientific Applications {\&} Novel Device Concepts}}}}\ (\bibinfo
  {publisher} {World Scientific},\ \bibinfo {year} {2003})\BibitemShut
  {NoStop}%
\bibitem [{\citenamefont {Pornsuwancharoen}\ \emph {et~al.}(2013)\citenamefont
  {Pornsuwancharoen}, \citenamefont {Tasakorn}, \citenamefont {Yupapin},\ and\
  \citenamefont {Chaiyasoonthorn}}]{Pornsuwancharoen2013}%
  \BibitemOpen
  \bibfield  {author} {\bibinfo {author} {\bibfnamefont {N.}~\bibnamefont
  {Pornsuwancharoen}}, \bibinfo {author} {\bibfnamefont {M.}~\bibnamefont
  {Tasakorn}}, \bibinfo {author} {\bibfnamefont {P.~P.}\ \bibnamefont
  {Yupapin}}, \ and\ \bibinfo {author} {\bibfnamefont {S.}~\bibnamefont
  {Chaiyasoonthorn}},\ }\href {\doibase 10.1016/j.ijleo.2011.12.010} {\bibfield
   {journal} {\bibinfo  {journal} {Optik}\ }\textbf {\bibinfo {volume} {124}},\
  \bibinfo {pages} {446} (\bibinfo {year} {2013})}\BibitemShut {NoStop}%
\bibitem [{\citenamefont {Stantchev}\ \emph {et~al.}(2017)\citenamefont
  {Stantchev}, \citenamefont {Phillips}, \citenamefont {Hobson}, \citenamefont
  {Hornett}, \citenamefont {Padgett},\ and\ \citenamefont
  {Hendry}}]{stantchev2017}%
  \BibitemOpen
  \bibfield  {author} {\bibinfo {author} {\bibfnamefont {R.~I.}\ \bibnamefont
  {Stantchev}}, \bibinfo {author} {\bibfnamefont {D.~B.}\ \bibnamefont
  {Phillips}}, \bibinfo {author} {\bibfnamefont {P.}~\bibnamefont {Hobson}},
  \bibinfo {author} {\bibfnamefont {S.~M.}\ \bibnamefont {Hornett}}, \bibinfo
  {author} {\bibfnamefont {M.~J.}\ \bibnamefont {Padgett}}, \ and\ \bibinfo
  {author} {\bibfnamefont {E.}~\bibnamefont {Hendry}},\ }\href {\doibase
  10.1364/OPTICA.4.000989} {\bibfield  {journal} {\bibinfo  {journal} {Optica}\
  }\textbf {\bibinfo {volume} {4}},\ \bibinfo {pages} {989} (\bibinfo {year}
  {2017})}\BibitemShut {NoStop}%
\bibitem [{\citenamefont {F{\"{u}}ser}\ and\ \citenamefont
  {Bieler}(2014)}]{Fuser2014}%
  \BibitemOpen
  \bibfield  {author} {\bibinfo {author} {\bibfnamefont {H.}~\bibnamefont
  {F{\"{u}}ser}}\ and\ \bibinfo {author} {\bibfnamefont {M.}~\bibnamefont
  {Bieler}},\ }\href {\doibase 10.1007/s10762-013-0038-8} {\bibfield  {journal}
  {\bibinfo  {journal} {Journal of Infrared, Millimeter, and Terahertz Waves}\
  }\textbf {\bibinfo {volume} {35}},\ \bibinfo {pages} {585} (\bibinfo {year}
  {2014})}\BibitemShut {NoStop}%
\bibitem [{\citenamefont {Barmes}\ \emph {et~al.}(2013)\citenamefont {Barmes},
  \citenamefont {Witte},\ and\ \citenamefont {Eikema}}]{Barmes2013}%
  \BibitemOpen
  \bibfield  {author} {\bibinfo {author} {\bibfnamefont {I.}~\bibnamefont
  {Barmes}}, \bibinfo {author} {\bibfnamefont {S.}~\bibnamefont {Witte}}, \
  and\ \bibinfo {author} {\bibfnamefont {K.~S.~E.}\ \bibnamefont {Eikema}},\
  }\href {\doibase 10.1038/nphoton.2012.299} {\bibfield  {journal} {\bibinfo
  {journal} {Nature Photonics}\ }\textbf {\bibinfo {volume} {7}},\ \bibinfo
  {pages} {38} (\bibinfo {year} {2013})}\BibitemShut {NoStop}%
\bibitem [{\citenamefont {Dai}\ \emph {et~al.}(2011)\citenamefont {Dai},
  \citenamefont {Liu},\ and\ \citenamefont {Zhang}}]{Dai2011}%
  \BibitemOpen
  \bibfield  {author} {\bibinfo {author} {\bibfnamefont {J.}~\bibnamefont
  {Dai}}, \bibinfo {author} {\bibfnamefont {J.}~\bibnamefont {Liu}}, \ and\
  \bibinfo {author} {\bibfnamefont {X.-C.}\ \bibnamefont {Zhang}},\ }\href
  {\doibase 10.1109/JSTQE.2010.2047007} {\bibfield  {journal} {\bibinfo
  {journal} {IEEE Journal of Selected Topics in Quantum Electronics}\ }\textbf
  {\bibinfo {volume} {17}},\ \bibinfo {pages} {183} (\bibinfo {year}
  {2011})}\BibitemShut {NoStop}%
\bibitem [{\citenamefont {Dai}\ and\ \citenamefont {Zhang}(2009)}]{Dai2009}%
  \BibitemOpen
  \bibfield  {author} {\bibinfo {author} {\bibfnamefont {J.}~\bibnamefont
  {Dai}}\ and\ \bibinfo {author} {\bibfnamefont {X.-C.}\ \bibnamefont
  {Zhang}},\ }\href {\doibase 10.1063/1.3068501} {\bibfield  {journal}
  {\bibinfo  {journal} {Applied Physics Letters}\ }\textbf {\bibinfo {volume}
  {94}},\ \bibinfo {pages} {021117} (\bibinfo {year} {2009})}\BibitemShut
  {NoStop}%
\bibitem [{\citenamefont {Burghoff}\ \emph {et~al.}(2014)\citenamefont
  {Burghoff}, \citenamefont {Kao}, \citenamefont {Han}, \citenamefont {Chan},
  \citenamefont {Cai}, \citenamefont {Yang}, \citenamefont {Hayton},
  \citenamefont {Gao}, \citenamefont {Reno},\ and\ \citenamefont
  {Hu}}]{Burghoff2014}%
  \BibitemOpen
  \bibfield  {author} {\bibinfo {author} {\bibfnamefont {D.}~\bibnamefont
  {Burghoff}}, \bibinfo {author} {\bibfnamefont {T.-y.}\ \bibnamefont {Kao}},
  \bibinfo {author} {\bibfnamefont {N.}~\bibnamefont {Han}}, \bibinfo {author}
  {\bibfnamefont {C.~W.~I.}\ \bibnamefont {Chan}}, \bibinfo {author}
  {\bibfnamefont {X.}~\bibnamefont {Cai}}, \bibinfo {author} {\bibfnamefont
  {Y.}~\bibnamefont {Yang}}, \bibinfo {author} {\bibfnamefont {D.~J.}\
  \bibnamefont {Hayton}}, \bibinfo {author} {\bibfnamefont {J.-r.}\
  \bibnamefont {Gao}}, \bibinfo {author} {\bibfnamefont {J.~L.}\ \bibnamefont
  {Reno}}, \ and\ \bibinfo {author} {\bibfnamefont {Q.}~\bibnamefont {Hu}},\
  }\href {\doibase 10.1038/nphoton.2014.85} {\bibfield  {journal} {\bibinfo
  {journal} {Nature Photonics}\ }\textbf {\bibinfo {volume} {8}},\ \bibinfo
  {pages} {462} (\bibinfo {year} {2014})}\BibitemShut {NoStop}%
\bibitem [{\citenamefont {Tan}\ \emph {et~al.}(2012)\citenamefont {Tan},
  \citenamefont {Huang}, \citenamefont {Liu}, \citenamefont {Xiong},\ and\
  \citenamefont {Fan}}]{Tan2012}%
  \BibitemOpen
  \bibfield  {author} {\bibinfo {author} {\bibfnamefont {P.}~\bibnamefont
  {Tan}}, \bibinfo {author} {\bibfnamefont {J.}~\bibnamefont {Huang}}, \bibinfo
  {author} {\bibfnamefont {K.}~\bibnamefont {Liu}}, \bibinfo {author}
  {\bibfnamefont {Y.}~\bibnamefont {Xiong}}, \ and\ \bibinfo {author}
  {\bibfnamefont {M.}~\bibnamefont {Fan}},\ }\href {\doibase
  10.1007/s11432-011-4515-1} {\bibfield  {journal} {\bibinfo  {journal}
  {Science China Information Sciences}\ }\textbf {\bibinfo {volume} {55}},\
  \bibinfo {pages} {1} (\bibinfo {year} {2012})}\BibitemShut {NoStop}%
\bibitem [{\citenamefont {Altares~Menendez}\ and\ \citenamefont
  {Maes}(2017)}]{AltaresMenendez2017a}%
  \BibitemOpen
  \bibfield  {author} {\bibinfo {author} {\bibfnamefont {G.}~\bibnamefont
  {Altares~Menendez}}\ and\ \bibinfo {author} {\bibfnamefont {B.}~\bibnamefont
  {Maes}},\ }\href {\doibase 10.1103/PhysRevB.95.144307} {\bibfield  {journal}
  {\bibinfo  {journal} {Physical Review B}\ }\textbf {\bibinfo {volume} {95}},\
  \bibinfo {pages} {144307} (\bibinfo {year} {2017})}\BibitemShut {NoStop}%
\bibitem [{\citenamefont {Otsuji}\ \emph {et~al.}(2013)\citenamefont {Otsuji},
  \citenamefont {Tombet}, \citenamefont {Satou}, \citenamefont {Ryzhii},\ and\
  \citenamefont {Ryzhii}}]{Outsuji2013}%
  \BibitemOpen
  \bibfield  {author} {\bibinfo {author} {\bibfnamefont {T.}~\bibnamefont
  {Otsuji}}, \bibinfo {author} {\bibfnamefont {S.~B.}\ \bibnamefont {Tombet}},
  \bibinfo {author} {\bibfnamefont {A.}~\bibnamefont {Satou}}, \bibinfo
  {author} {\bibfnamefont {M.}~\bibnamefont {Ryzhii}}, \ and\ \bibinfo {author}
  {\bibfnamefont {V.}~\bibnamefont {Ryzhii}},\ }\href {\doibase
  10.1109/JSTQE.2012.2208734} {\bibfield  {journal} {\bibinfo  {journal} {IEEE
  Journal of Selected Topics in Quantum Electronics}\ }\textbf {\bibinfo
  {volume} {19}},\ \bibinfo {pages} {8400209} (\bibinfo {year}
  {2013})}\BibitemShut {NoStop}%
\bibitem [{\citenamefont {Otsuji}\ \emph {et~al.}(2014)\citenamefont {Otsuji},
  \citenamefont {Satou}, \citenamefont {Ryzhii}, \citenamefont {Popov},
  \citenamefont {Mitin},\ and\ \citenamefont {Ryzhii}}]{Otsuji2014b}%
  \BibitemOpen
  \bibfield  {author} {\bibinfo {author} {\bibfnamefont {T.}~\bibnamefont
  {Otsuji}}, \bibinfo {author} {\bibfnamefont {A.}~\bibnamefont {Satou}},
  \bibinfo {author} {\bibfnamefont {M.}~\bibnamefont {Ryzhii}}, \bibinfo
  {author} {\bibfnamefont {V.}~\bibnamefont {Popov}}, \bibinfo {author}
  {\bibfnamefont {V.}~\bibnamefont {Mitin}}, \ and\ \bibinfo {author}
  {\bibfnamefont {V.}~\bibnamefont {Ryzhii}},\ }\href {\doibase
  10.1088/1742-6596/486/1/012007} {\bibfield  {journal} {\bibinfo  {journal}
  {Journal of Physics: Conference Series}\ }\textbf {\bibinfo {volume} {486}},\
  \bibinfo {pages} {012007} (\bibinfo {year} {2014})}\BibitemShut {NoStop}%
\bibitem [{\citenamefont {Schwierz}(2010)}]{Schwierz2010}%
  \BibitemOpen
  \bibfield  {author} {\bibinfo {author} {\bibfnamefont {F.}~\bibnamefont
  {Schwierz}},\ }\href {\doibase 10.1038/nnano.2010.89} {\bibfield  {journal}
  {\bibinfo  {journal} {Nature Nanotechnology}\ }\textbf {\bibinfo {volume}
  {5}},\ \bibinfo {pages} {487} (\bibinfo {year} {2010})}\BibitemShut {NoStop}%
\bibitem [{\citenamefont {Kim}\ \emph {et~al.}(2015)\citenamefont {Kim},
  \citenamefont {Kim}, \citenamefont {Cho}, \citenamefont {Ryoo}, \citenamefont
  {Park}, \citenamefont {Kim}, \citenamefont {Kim}, \citenamefont {Lee},
  \citenamefont {Li}, \citenamefont {Park}, \citenamefont {Shim~Yoo},
  \citenamefont {Yoon}, \citenamefont {Dorgan}, \citenamefont {Pop},
  \citenamefont {Heinz}, \citenamefont {Hone}, \citenamefont {Chun},
  \citenamefont {Cheong}, \citenamefont {Lee}, \citenamefont {Bae},\ and\
  \citenamefont {Park}}]{Kim2015}%
  \BibitemOpen
  \bibfield  {author} {\bibinfo {author} {\bibfnamefont {Y.~S. Y.~D.}\
  \bibnamefont {Kim}}, \bibinfo {author} {\bibfnamefont {H.}~\bibnamefont
  {Kim}}, \bibinfo {author} {\bibfnamefont {Y.}~\bibnamefont {Cho}}, \bibinfo
  {author} {\bibfnamefont {J.~H.}\ \bibnamefont {Ryoo}}, \bibinfo {author}
  {\bibfnamefont {C.-H.}\ \bibnamefont {Park}}, \bibinfo {author}
  {\bibfnamefont {P.}~\bibnamefont {Kim}}, \bibinfo {author} {\bibfnamefont
  {Y.~S. Y.~D.}\ \bibnamefont {Kim}}, \bibinfo {author} {\bibfnamefont
  {S.~S.~W.}\ \bibnamefont {Lee}}, \bibinfo {author} {\bibfnamefont
  {Y.}~\bibnamefont {Li}}, \bibinfo {author} {\bibfnamefont {S.-N.}\
  \bibnamefont {Park}}, \bibinfo {author} {\bibfnamefont {Y.}~\bibnamefont
  {Shim~Yoo}}, \bibinfo {author} {\bibfnamefont {D.}~\bibnamefont {Yoon}},
  \bibinfo {author} {\bibfnamefont {V.~E.}\ \bibnamefont {Dorgan}}, \bibinfo
  {author} {\bibfnamefont {E.}~\bibnamefont {Pop}}, \bibinfo {author}
  {\bibfnamefont {T.~F.}\ \bibnamefont {Heinz}}, \bibinfo {author}
  {\bibfnamefont {J.}~\bibnamefont {Hone}}, \bibinfo {author} {\bibfnamefont
  {S.-H.}\ \bibnamefont {Chun}}, \bibinfo {author} {\bibfnamefont
  {H.}~\bibnamefont {Cheong}}, \bibinfo {author} {\bibfnamefont {S.~S.~W.}\
  \bibnamefont {Lee}}, \bibinfo {author} {\bibfnamefont {M.-H.}\ \bibnamefont
  {Bae}}, \ and\ \bibinfo {author} {\bibfnamefont {Y.~D.}\ \bibnamefont
  {Park}},\ }\href {\doibase 10.1038/nnano.2015.118} {\bibfield  {journal}
  {\bibinfo  {journal} {Nature Nanotechnology}\ }\textbf {\bibinfo {volume}
  {10}},\ \bibinfo {pages} {676} (\bibinfo {year} {2015})}\BibitemShut
  {NoStop}%
\bibitem [{\citenamefont {Kim}\ \emph {et~al.}(2018)\citenamefont {Kim},
  \citenamefont {Gao}, \citenamefont {Shiue}, \citenamefont {Wang},
  \citenamefont {Aslan}, \citenamefont {Bae}, \citenamefont {Kim},
  \citenamefont {Seo}, \citenamefont {Choi}, \citenamefont {Kim}, \citenamefont
  {Nemilentsau}, \citenamefont {Low}, \citenamefont {Tan}, \citenamefont
  {Efetov}, \citenamefont {Taniguchi}, \citenamefont {Watanabe}, \citenamefont
  {Shepard}, \citenamefont {Heinz}, \citenamefont {Englund}, \citenamefont
  {Hone}, \citenamefont {Kim}, \citenamefont {Choi}, \citenamefont {Gao},
  \citenamefont {Aslan}, \citenamefont {Wang}, \citenamefont {Kim},
  \citenamefont {Heinz}, \citenamefont {Hone}, \citenamefont {Bae},
  \citenamefont {Kim}, \citenamefont {Efetov}, \citenamefont {Shepard},
  \citenamefont {Englund}, \citenamefont {Taniguchi}, \citenamefont
  {Nemilentsau}, \citenamefont {Low}, \citenamefont {Shiue}, \citenamefont
  {Wang}, \citenamefont {Aslan}, \citenamefont {Bae}, \citenamefont {Kim},
  \citenamefont {Seo}, \citenamefont {Choi}, \citenamefont {Kim}, \citenamefont
  {Nemilentsau}, \citenamefont {Low}, \citenamefont {Tan}, \citenamefont
  {Efetov}, \citenamefont {Taniguchi}, \citenamefont {Watanabe}, \citenamefont
  {Shepard}, \citenamefont {Heinz}, \citenamefont {Englund},\ and\
  \citenamefont {Hone}}]{Kim2018}%
  \BibitemOpen
  \bibfield  {author} {\bibinfo {author} {\bibfnamefont {Y.~D.}\ \bibnamefont
  {Kim}}, \bibinfo {author} {\bibfnamefont {Y.}~\bibnamefont {Gao}}, \bibinfo
  {author} {\bibfnamefont {R.-J.}\ \bibnamefont {Shiue}}, \bibinfo {author}
  {\bibfnamefont {L.}~\bibnamefont {Wang}}, \bibinfo {author} {\bibfnamefont
  {O.~B.}\ \bibnamefont {Aslan}}, \bibinfo {author} {\bibfnamefont {M.-H.}\
  \bibnamefont {Bae}}, \bibinfo {author} {\bibfnamefont {H.}~\bibnamefont
  {Kim}}, \bibinfo {author} {\bibfnamefont {D.}~\bibnamefont {Seo}}, \bibinfo
  {author} {\bibfnamefont {H.-J.}\ \bibnamefont {Choi}}, \bibinfo {author}
  {\bibfnamefont {S.~H.}\ \bibnamefont {Kim}}, \bibinfo {author} {\bibfnamefont
  {A.}~\bibnamefont {Nemilentsau}}, \bibinfo {author} {\bibfnamefont
  {T.}~\bibnamefont {Low}}, \bibinfo {author} {\bibfnamefont {C.}~\bibnamefont
  {Tan}}, \bibinfo {author} {\bibfnamefont {D.~K.}\ \bibnamefont {Efetov}},
  \bibinfo {author} {\bibfnamefont {T.}~\bibnamefont {Taniguchi}}, \bibinfo
  {author} {\bibfnamefont {K.}~\bibnamefont {Watanabe}}, \bibinfo {author}
  {\bibfnamefont {K.~L.}\ \bibnamefont {Shepard}}, \bibinfo {author}
  {\bibfnamefont {T.~F.}\ \bibnamefont {Heinz}}, \bibinfo {author}
  {\bibfnamefont {D.}~\bibnamefont {Englund}}, \bibinfo {author} {\bibfnamefont
  {J.}~\bibnamefont {Hone}}, \bibinfo {author} {\bibfnamefont {Y.~D.}\
  \bibnamefont {Kim}}, \bibinfo {author} {\bibfnamefont {H.-J.}\ \bibnamefont
  {Choi}}, \bibinfo {author} {\bibfnamefont {Y.}~\bibnamefont {Gao}}, \bibinfo
  {author} {\bibfnamefont {O.~B.}\ \bibnamefont {Aslan}}, \bibinfo {author}
  {\bibfnamefont {L.}~\bibnamefont {Wang}}, \bibinfo {author} {\bibfnamefont
  {H.}~\bibnamefont {Kim}}, \bibinfo {author} {\bibfnamefont {T.~F.}\
  \bibnamefont {Heinz}}, \bibinfo {author} {\bibfnamefont {J.}~\bibnamefont
  {Hone}}, \bibinfo {author} {\bibfnamefont {M.-H.}\ \bibnamefont {Bae}},
  \bibinfo {author} {\bibfnamefont {S.~H.}\ \bibnamefont {Kim}}, \bibinfo
  {author} {\bibfnamefont {D.~K.}\ \bibnamefont {Efetov}}, \bibinfo {author}
  {\bibfnamefont {K.~L.}\ \bibnamefont {Shepard}}, \bibinfo {author}
  {\bibfnamefont {D.}~\bibnamefont {Englund}}, \bibinfo {author} {\bibfnamefont
  {T.}~\bibnamefont {Taniguchi}}, \bibinfo {author} {\bibfnamefont
  {A.}~\bibnamefont {Nemilentsau}}, \bibinfo {author} {\bibfnamefont
  {T.}~\bibnamefont {Low}}, \bibinfo {author} {\bibfnamefont {R.-J.}\
  \bibnamefont {Shiue}}, \bibinfo {author} {\bibfnamefont {L.}~\bibnamefont
  {Wang}}, \bibinfo {author} {\bibfnamefont {O.~B.}\ \bibnamefont {Aslan}},
  \bibinfo {author} {\bibfnamefont {M.-H.}\ \bibnamefont {Bae}}, \bibinfo
  {author} {\bibfnamefont {H.}~\bibnamefont {Kim}}, \bibinfo {author}
  {\bibfnamefont {D.}~\bibnamefont {Seo}}, \bibinfo {author} {\bibfnamefont
  {H.-J.}\ \bibnamefont {Choi}}, \bibinfo {author} {\bibfnamefont {S.~H.}\
  \bibnamefont {Kim}}, \bibinfo {author} {\bibfnamefont {A.}~\bibnamefont
  {Nemilentsau}}, \bibinfo {author} {\bibfnamefont {T.}~\bibnamefont {Low}},
  \bibinfo {author} {\bibfnamefont {C.}~\bibnamefont {Tan}}, \bibinfo {author}
  {\bibfnamefont {D.~K.}\ \bibnamefont {Efetov}}, \bibinfo {author}
  {\bibfnamefont {T.}~\bibnamefont {Taniguchi}}, \bibinfo {author}
  {\bibfnamefont {K.}~\bibnamefont {Watanabe}}, \bibinfo {author}
  {\bibfnamefont {K.~L.}\ \bibnamefont {Shepard}}, \bibinfo {author}
  {\bibfnamefont {T.~F.}\ \bibnamefont {Heinz}}, \bibinfo {author}
  {\bibfnamefont {D.}~\bibnamefont {Englund}}, \ and\ \bibinfo {author}
  {\bibfnamefont {J.}~\bibnamefont {Hone}},\ }\href {\doibase
  10.1021/acs.nanolett.7b04324} {\bibfield  {journal} {\bibinfo  {journal}
  {Nano Letters}\ }\textbf {\bibinfo {volume} {18}},\ \bibinfo {pages} {934}
  (\bibinfo {year} {2018})}\BibitemShut {NoStop}%
\bibitem [{\citenamefont {Beltaos}\ \emph {et~al.}(2017)\citenamefont
  {Beltaos}, \citenamefont {Bergren}, \citenamefont {Bosnick}, \citenamefont
  {Pekas}, \citenamefont {Lane}, \citenamefont {Cui}, \citenamefont
  {Matkovi{\'{c}}}, \citenamefont {Meldrum}, \citenamefont {Bosnick},
  \citenamefont {Cui}, \citenamefont {Bergren}, \citenamefont {Beltaos},
  \citenamefont {Lane}, \citenamefont {Matkovi{\'{c}}}, \citenamefont
  {Bergren}, \citenamefont {Bosnick}, \citenamefont {Pekas}, \citenamefont
  {Lane}, \citenamefont {Cui}, \citenamefont {Matkovi{\'{c}}},\ and\
  \citenamefont {Meldrum}}]{Beltaos2017}%
  \BibitemOpen
  \bibfield  {author} {\bibinfo {author} {\bibfnamefont {A.}~\bibnamefont
  {Beltaos}}, \bibinfo {author} {\bibfnamefont {A.~J.}\ \bibnamefont
  {Bergren}}, \bibinfo {author} {\bibfnamefont {K.}~\bibnamefont {Bosnick}},
  \bibinfo {author} {\bibfnamefont {N.}~\bibnamefont {Pekas}}, \bibinfo
  {author} {\bibfnamefont {S.}~\bibnamefont {Lane}}, \bibinfo {author}
  {\bibfnamefont {K.}~\bibnamefont {Cui}}, \bibinfo {author} {\bibfnamefont
  {A.}~\bibnamefont {Matkovi{\'{c}}}}, \bibinfo {author} {\bibfnamefont
  {A.}~\bibnamefont {Meldrum}}, \bibinfo {author} {\bibfnamefont
  {K.}~\bibnamefont {Bosnick}}, \bibinfo {author} {\bibfnamefont
  {K.}~\bibnamefont {Cui}}, \bibinfo {author} {\bibfnamefont {A.~J.}\
  \bibnamefont {Bergren}}, \bibinfo {author} {\bibfnamefont {A.}~\bibnamefont
  {Beltaos}}, \bibinfo {author} {\bibfnamefont {S.}~\bibnamefont {Lane}},
  \bibinfo {author} {\bibfnamefont {A.}~\bibnamefont {Matkovi{\'{c}}}},
  \bibinfo {author} {\bibfnamefont {A.~J.}\ \bibnamefont {Bergren}}, \bibinfo
  {author} {\bibfnamefont {K.}~\bibnamefont {Bosnick}}, \bibinfo {author}
  {\bibfnamefont {N.}~\bibnamefont {Pekas}}, \bibinfo {author} {\bibfnamefont
  {S.}~\bibnamefont {Lane}}, \bibinfo {author} {\bibfnamefont {K.}~\bibnamefont
  {Cui}}, \bibinfo {author} {\bibfnamefont {A.}~\bibnamefont {Matkovi{\'{c}}}},
  \ and\ \bibinfo {author} {\bibfnamefont {A.}~\bibnamefont {Meldrum}},\ }\href
  {\doibase 10.1088/2399-1984/aa8b04} {\bibfield  {journal} {\bibinfo
  {journal} {Nano Futures}\ }\textbf {\bibinfo {volume} {1}},\ \bibinfo {pages}
  {025004} (\bibinfo {year} {2017})}\BibitemShut {NoStop}%
\bibitem [{\citenamefont {Yadav}\ \emph {et~al.}(2017)\citenamefont {Yadav},
  \citenamefont {Tobah}, \citenamefont {Mitsushio}, \citenamefont {Tamamushi},
  \citenamefont {Watanabe}, \citenamefont {Dubinov}, \citenamefont {Ryzhii},
  \citenamefont {Ryzhii},\ and\ \citenamefont {Otsuji}}]{Yadav2017a}%
  \BibitemOpen
  \bibfield  {author} {\bibinfo {author} {\bibfnamefont {D.}~\bibnamefont
  {Yadav}}, \bibinfo {author} {\bibfnamefont {Y.}~\bibnamefont {Tobah}},
  \bibinfo {author} {\bibfnamefont {J.}~\bibnamefont {Mitsushio}}, \bibinfo
  {author} {\bibfnamefont {G.}~\bibnamefont {Tamamushi}}, \bibinfo {author}
  {\bibfnamefont {T.}~\bibnamefont {Watanabe}}, \bibinfo {author}
  {\bibfnamefont {A.~A.}\ \bibnamefont {Dubinov}}, \bibinfo {author}
  {\bibfnamefont {M.}~\bibnamefont {Ryzhii}}, \bibinfo {author} {\bibfnamefont
  {V.}~\bibnamefont {Ryzhii}}, \ and\ \bibinfo {author} {\bibfnamefont
  {T.}~\bibnamefont {Otsuji}},\ }in\ \href {\doibase
  10.1364/CLEO{\_}AT.2017.AM2B.7} {\emph {\bibinfo {booktitle} {Conference on
  Lasers and Electro-Optics}}}\ (\bibinfo  {publisher} {OSA},\ \bibinfo
  {address} {Washington, D.C.},\ \bibinfo {year} {2017})\ p.\ \bibinfo {pages}
  {AM2B.7}\BibitemShut {NoStop}%
\bibitem [{\citenamefont {Yadav}\ \emph {et~al.}(2018)\citenamefont {Yadav},
  \citenamefont {Tamamushi}, \citenamefont {Watanabe}, \citenamefont
  {Mitsushio}, \citenamefont {Tobah}, \citenamefont {Sugawara}, \citenamefont
  {Dubinov}, \citenamefont {Satou}, \citenamefont {Ryzhii}, \citenamefont
  {Ryzhii},\ and\ \citenamefont {Otsuji}}]{Yadav2018a}%
  \BibitemOpen
  \bibfield  {author} {\bibinfo {author} {\bibfnamefont {D.}~\bibnamefont
  {Yadav}}, \bibinfo {author} {\bibfnamefont {G.}~\bibnamefont {Tamamushi}},
  \bibinfo {author} {\bibfnamefont {T.}~\bibnamefont {Watanabe}}, \bibinfo
  {author} {\bibfnamefont {J.}~\bibnamefont {Mitsushio}}, \bibinfo {author}
  {\bibfnamefont {Y.}~\bibnamefont {Tobah}}, \bibinfo {author} {\bibfnamefont
  {K.}~\bibnamefont {Sugawara}}, \bibinfo {author} {\bibfnamefont {A.~A.}\
  \bibnamefont {Dubinov}}, \bibinfo {author} {\bibfnamefont {A.}~\bibnamefont
  {Satou}}, \bibinfo {author} {\bibfnamefont {M.}~\bibnamefont {Ryzhii}},
  \bibinfo {author} {\bibfnamefont {V.}~\bibnamefont {Ryzhii}}, \ and\ \bibinfo
  {author} {\bibfnamefont {T.}~\bibnamefont {Otsuji}},\ }\href {\doibase
  10.1515/nanoph-2017-0106} {\bibfield  {journal} {\bibinfo  {journal}
  {Nanophotonics}\ }\textbf {\bibinfo {volume} {7}},\ \bibinfo {pages} {741}
  (\bibinfo {year} {2018})}\BibitemShut {NoStop}%
\bibitem [{\citenamefont {Dyakonov}\ and\ \citenamefont
  {Shur}(1993)}]{Dyakonov1993}%
  \BibitemOpen
  \bibfield  {author} {\bibinfo {author} {\bibfnamefont {M.}~\bibnamefont
  {Dyakonov}}\ and\ \bibinfo {author} {\bibfnamefont {M.}~\bibnamefont
  {Shur}},\ }\href {\doibase 10.1103/PhysRevLett.71.2465} {\bibfield  {journal}
  {\bibinfo  {journal} {Physical Review Letters}\ }\textbf {\bibinfo {volume}
  {71}},\ \bibinfo {pages} {2465} (\bibinfo {year} {1993})}\BibitemShut
  {NoStop}%
\bibitem [{\citenamefont {Kargar}\ \emph {et~al.}(2018)\citenamefont {Kargar},
  \citenamefont {Linn},\ and\ \citenamefont {Jungemann}}]{Kargar2018}%
  \BibitemOpen
  \bibfield  {author} {\bibinfo {author} {\bibfnamefont {Z.}~\bibnamefont
  {Kargar}}, \bibinfo {author} {\bibfnamefont {T.}~\bibnamefont {Linn}}, \ and\
  \bibinfo {author} {\bibfnamefont {C.}~\bibnamefont {Jungemann}},\ }\href@noop
  {} {\bibfield  {journal} {\bibinfo  {journal} {Semiconductor Science and
  Technology}\ }\textbf {\bibinfo {volume} {33}} (\bibinfo {year}
  {2018})}\BibitemShut {NoStop}%
\bibitem [{\citenamefont {Chaves}\ \emph {et~al.}(2017)\citenamefont {Chaves},
  \citenamefont {Peres}, \citenamefont {Smirnov},\ and\ \citenamefont
  {Asger~Mortensen}}]{Chaves2017}%
  \BibitemOpen
  \bibfield  {author} {\bibinfo {author} {\bibfnamefont {A.~J.}\ \bibnamefont
  {Chaves}}, \bibinfo {author} {\bibfnamefont {N.~M.}\ \bibnamefont {Peres}},
  \bibinfo {author} {\bibfnamefont {G.}~\bibnamefont {Smirnov}}, \ and\
  \bibinfo {author} {\bibfnamefont {N.}~\bibnamefont {Asger~Mortensen}},\
  }\href {\doibase 10.1103/PhysRevB.96.195438} {\bibfield  {journal} {\bibinfo
  {journal} {Physical Review B}\ }\textbf {\bibinfo {volume} {96}},\ \bibinfo
  {pages} {1} (\bibinfo {year} {2017})}\BibitemShut {NoStop}%
\bibitem [{\citenamefont {M{\"{u}}ller}\ \emph {et~al.}(2009)\citenamefont
  {M{\"{u}}ller}, \citenamefont {Schmalian},\ and\ \citenamefont
  {Fritz}}]{Muller2009}%
  \BibitemOpen
  \bibfield  {author} {\bibinfo {author} {\bibfnamefont {M.}~\bibnamefont
  {M{\"{u}}ller}}, \bibinfo {author} {\bibfnamefont {J.}~\bibnamefont
  {Schmalian}}, \ and\ \bibinfo {author} {\bibfnamefont {L.}~\bibnamefont
  {Fritz}},\ }\href {\doibase 10.1103/PhysRevLett.103.025301} {\bibfield
  {journal} {\bibinfo  {journal} {Physical Review Letters}\ }\textbf {\bibinfo
  {volume} {103}},\ \bibinfo {pages} {025301} (\bibinfo {year}
  {2009})}\BibitemShut {NoStop}%
\bibitem [{\citenamefont {Svintsov}\ \emph {et~al.}(2013)\citenamefont
  {Svintsov}, \citenamefont {Vyurkov}, \citenamefont {Ryzhii},\ and\
  \citenamefont {Otsuji}}]{Svintsov2013}%
  \BibitemOpen
  \bibfield  {author} {\bibinfo {author} {\bibfnamefont {D.}~\bibnamefont
  {Svintsov}}, \bibinfo {author} {\bibfnamefont {V.}~\bibnamefont {Vyurkov}},
  \bibinfo {author} {\bibfnamefont {V.}~\bibnamefont {Ryzhii}}, \ and\ \bibinfo
  {author} {\bibfnamefont {T.}~\bibnamefont {Otsuji}},\ }\href {\doibase
  10.1103/PhysRevB.88.245444} {\bibfield  {journal} {\bibinfo  {journal}
  {Physical Review B - Condensed Matter and Materials Physics}\ }\textbf
  {\bibinfo {volume} {88}},\ \bibinfo {pages} {28} (\bibinfo {year}
  {2013})}\BibitemShut {NoStop}%
\bibitem [{\citenamefont {Castro~Neto}\ \emph {et~al.}(2009)\citenamefont
  {Castro~Neto}, \citenamefont {Guinea}, \citenamefont {Peres}, \citenamefont
  {Novoselov},\ and\ \citenamefont {Geim}}]{Neto2007}%
  \BibitemOpen
  \bibfield  {author} {\bibinfo {author} {\bibfnamefont {A.~H.}\ \bibnamefont
  {Castro~Neto}}, \bibinfo {author} {\bibfnamefont {F.}~\bibnamefont {Guinea}},
  \bibinfo {author} {\bibfnamefont {N.~M.~R.}\ \bibnamefont {Peres}}, \bibinfo
  {author} {\bibfnamefont {K.~S.}\ \bibnamefont {Novoselov}}, \ and\ \bibinfo
  {author} {\bibfnamefont {A.~K.}\ \bibnamefont {Geim}},\ }\href {\doibase
  10.1103/RevModPhys.81.109} {\bibfield  {journal} {\bibinfo  {journal}
  {Reviews of Modern Physics}\ }\textbf {\bibinfo {volume} {81}},\ \bibinfo
  {pages} {109} (\bibinfo {year} {2009})}\BibitemShut {NoStop}%
\bibitem [{\citenamefont {Zhu}\ \emph {et~al.}(2009)\citenamefont {Zhu},
  \citenamefont {Perebeinos}, \citenamefont {Freitag},\ and\ \citenamefont
  {Avouris}}]{Zhu2009}%
  \BibitemOpen
  \bibfield  {author} {\bibinfo {author} {\bibfnamefont {W.}~\bibnamefont
  {Zhu}}, \bibinfo {author} {\bibfnamefont {V.}~\bibnamefont {Perebeinos}},
  \bibinfo {author} {\bibfnamefont {M.}~\bibnamefont {Freitag}}, \ and\
  \bibinfo {author} {\bibfnamefont {P.}~\bibnamefont {Avouris}},\ }\href
  {\doibase 10.1103/PhysRevB.80.235402} {\bibfield  {journal} {\bibinfo
  {journal} {Physical Review B}\ }\textbf {\bibinfo {volume} {80}},\ \bibinfo
  {pages} {235402} (\bibinfo {year} {2009})}\BibitemShut {NoStop}%
\bibitem [{\citenamefont {Fang}\ \emph {et~al.}(2007)\citenamefont {Fang},
  \citenamefont {Konar}, \citenamefont {Xing},\ and\ \citenamefont
  {Jena}}]{Fang2007}%
  \BibitemOpen
  \bibfield  {author} {\bibinfo {author} {\bibfnamefont {T.}~\bibnamefont
  {Fang}}, \bibinfo {author} {\bibfnamefont {A.}~\bibnamefont {Konar}},
  \bibinfo {author} {\bibfnamefont {H.}~\bibnamefont {Xing}}, \ and\ \bibinfo
  {author} {\bibfnamefont {D.}~\bibnamefont {Jena}},\ }\href {\doibase
  10.1063/1.2776887} {\bibfield  {journal} {\bibinfo  {journal} {Applied
  Physics Letters}\ }\textbf {\bibinfo {volume} {91}},\ \bibinfo {pages} {2007}
  (\bibinfo {year} {2007})}\BibitemShut {NoStop}%
\bibitem [{\citenamefont {Dr{\"{o}}scher}\ \emph {et~al.}(2012)\citenamefont
  {Dr{\"{o}}scher}, \citenamefont {Roulleau}, \citenamefont {Molitor},
  \citenamefont {Studerus}, \citenamefont {Stampfer}, \citenamefont {Ensslin},\
  and\ \citenamefont {Ihn}}]{Droscher2012}%
  \BibitemOpen
  \bibfield  {author} {\bibinfo {author} {\bibfnamefont {S.}~\bibnamefont
  {Dr{\"{o}}scher}}, \bibinfo {author} {\bibfnamefont {P.}~\bibnamefont
  {Roulleau}}, \bibinfo {author} {\bibfnamefont {F.}~\bibnamefont {Molitor}},
  \bibinfo {author} {\bibfnamefont {P.}~\bibnamefont {Studerus}}, \bibinfo
  {author} {\bibfnamefont {C.}~\bibnamefont {Stampfer}}, \bibinfo {author}
  {\bibfnamefont {K.}~\bibnamefont {Ensslin}}, \ and\ \bibinfo {author}
  {\bibfnamefont {T.}~\bibnamefont {Ihn}},\ }\href {\doibase
  10.1088/0031-8949/2012/T146/014009} {\bibfield  {journal} {\bibinfo
  {journal} {Physica Scripta}\ }\textbf {\bibinfo {volume} {T146}},\ \bibinfo
  {pages} {014009} (\bibinfo {year} {2012})}\BibitemShut {NoStop}%
\bibitem [{\citenamefont {Das~Sarma}\ \emph {et~al.}(2011)\citenamefont
  {Das~Sarma}, \citenamefont {Adam}, \citenamefont {Hwang},\ and\ \citenamefont
  {Rossi}}]{Sarma2010}%
  \BibitemOpen
  \bibfield  {author} {\bibinfo {author} {\bibfnamefont {S.}~\bibnamefont
  {Das~Sarma}}, \bibinfo {author} {\bibfnamefont {S.}~\bibnamefont {Adam}},
  \bibinfo {author} {\bibfnamefont {E.~H.}\ \bibnamefont {Hwang}}, \ and\
  \bibinfo {author} {\bibfnamefont {E.}~\bibnamefont {Rossi}},\ }\href
  {\doibase 10.1103/RevModPhys.83.407} {\bibfield  {journal} {\bibinfo
  {journal} {Reviews of Modern Physics}\ }\textbf {\bibinfo {volume} {83}},\
  \bibinfo {pages} {407} (\bibinfo {year} {2011})}\BibitemShut {NoStop}%
\bibitem [{\citenamefont {Dyakonov}(2011)}]{Dyakonov2011}%
  \BibitemOpen
  \bibfield  {author} {\bibinfo {author} {\bibfnamefont {M.}~\bibnamefont
  {Dyakonov}},\ }\href {\doibase 10.1016/j.crhy.2010.05.003} {\bibfield
  {journal} {\bibinfo  {journal} {Comptes Rendus Physique}\ }\textbf {\bibinfo
  {volume} {11}},\ \bibinfo {pages} {10} (\bibinfo {year} {2011})}\BibitemShut
  {NoStop}%
\bibitem [{\citenamefont {Dmitriev}\ \emph {et~al.}(1997)\citenamefont
  {Dmitriev}, \citenamefont {Furman}, \citenamefont {Kachorovskii},
  \citenamefont {Samsonidze},\ and\ \citenamefont {Samsonidze}}]{Dmitriev1997}%
  \BibitemOpen
  \bibfield  {author} {\bibinfo {author} {\bibfnamefont {A.~P.}\ \bibnamefont
  {Dmitriev}}, \bibinfo {author} {\bibfnamefont {A.~S.}\ \bibnamefont
  {Furman}}, \bibinfo {author} {\bibfnamefont {V.~Y.}\ \bibnamefont
  {Kachorovskii}}, \bibinfo {author} {\bibfnamefont {G.~G.}\ \bibnamefont
  {Samsonidze}}, \ and\ \bibinfo {author} {\bibfnamefont {G.~G.}\ \bibnamefont
  {Samsonidze}},\ }\href@noop {} {\bibfield  {journal} {\bibinfo  {journal}
  {Physical Review B}\ }\textbf {\bibinfo {volume} {55}},\ \bibinfo {pages}
  {10319} (\bibinfo {year} {1997})}\BibitemShut {NoStop}%
\bibitem [{\citenamefont {Crowne}(1997)}]{Crowne1997}%
  \BibitemOpen
  \bibfield  {author} {\bibinfo {author} {\bibfnamefont {F.~J.}\ \bibnamefont
  {Crowne}},\ }\href {\doibase 10.1063/1.365895} {\bibfield  {journal}
  {\bibinfo  {journal} {Journal of Applied Physics}\ }\textbf {\bibinfo
  {volume} {82}},\ \bibinfo {pages} {1242} (\bibinfo {year}
  {1997})}\BibitemShut {NoStop}%
\bibitem [{\citenamefont {Bolotin}\ \emph {et~al.}(2008)\citenamefont
  {Bolotin}, \citenamefont {Sikes}, \citenamefont {Jiang}, \citenamefont
  {Klima}, \citenamefont {Fudenberg}, \citenamefont {Hone}, \citenamefont
  {Kim},\ and\ \citenamefont {Stormer}}]{Bolotin2008}%
  \BibitemOpen
  \bibfield  {author} {\bibinfo {author} {\bibfnamefont {K.~I.}\ \bibnamefont
  {Bolotin}}, \bibinfo {author} {\bibfnamefont {K.~J.}\ \bibnamefont {Sikes}},
  \bibinfo {author} {\bibfnamefont {Z.}~\bibnamefont {Jiang}}, \bibinfo
  {author} {\bibfnamefont {M.}~\bibnamefont {Klima}}, \bibinfo {author}
  {\bibfnamefont {G.}~\bibnamefont {Fudenberg}}, \bibinfo {author}
  {\bibfnamefont {J.}~\bibnamefont {Hone}}, \bibinfo {author} {\bibfnamefont
  {P.}~\bibnamefont {Kim}}, \ and\ \bibinfo {author} {\bibfnamefont {H.~L.}\
  \bibnamefont {Stormer}},\ }\href {\doibase 10.1016/j.ssc.2008.02.024}
  {\bibfield  {journal} {\bibinfo  {journal} {Solid State Communications}\
  }\textbf {\bibinfo {volume} {146}},\ \bibinfo {pages} {351} (\bibinfo {year}
  {2008})}\BibitemShut {NoStop}%
\bibitem [{\citenamefont {LeVeque}(1992)}]{LeVeque1992}%
  \BibitemOpen
  \bibfield  {author} {\bibinfo {author} {\bibfnamefont {R.~J.}\ \bibnamefont
  {LeVeque}},\ }\href@noop {} {\emph {\bibinfo {title} {Lectures in Mathematics
  ETH Z{\"{u}}rich}}},\ \bibinfo {edition} {2nd}\ ed.\ (\bibinfo  {publisher}
  {Birkhauser Verlag},\ \bibinfo {address} {Boston},\ \bibinfo {year}
  {1992})\BibitemShut {NoStop}%
\bibitem [{\citenamefont {Griffiths}\ and\ \citenamefont
  {Heald}(1991)}]{Griffiths1991}%
  \BibitemOpen
  \bibfield  {author} {\bibinfo {author} {\bibfnamefont {D.~J.}\ \bibnamefont
  {Griffiths}}\ and\ \bibinfo {author} {\bibfnamefont {M.~A.}\ \bibnamefont
  {Heald}},\ }\href {\doibase 10.1119/1.16589} {\bibfield  {journal} {\bibinfo
  {journal} {American Journal of Physics}\ }\textbf {\bibinfo {volume} {59}},\
  \bibinfo {pages} {111} (\bibinfo {year} {1991})}\BibitemShut {NoStop}%
\bibitem [{\citenamefont {Kobayashi}(2016)}]{Kobayashi2016}%
  \BibitemOpen
  \bibfield  {author} {\bibinfo {author} {\bibfnamefont {H.}~\bibnamefont
  {Kobayashi}},\ }in\ \href {\doibase 10.1109/ursi-emts.2016.7571501} {\emph
  {\bibinfo {booktitle} {2016 {URSI} International Symposium on Electromagnetic
  Theory ({EMTS})}}}\ (\bibinfo  {publisher} {{IEEE}},\ \bibinfo {year}
  {2016})\BibitemShut {NoStop}%
\bibitem [{\citenamefont {Frigo}\ and\ \citenamefont
  {Johnson}(2005)}]{Frigo2005}%
  \BibitemOpen
  \bibfield  {author} {\bibinfo {author} {\bibfnamefont {M.}~\bibnamefont
  {Frigo}}\ and\ \bibinfo {author} {\bibfnamefont {S.}~\bibnamefont
  {Johnson}},\ }\href {\doibase 10.1109/JPROC.2004.840301} {\bibfield
  {journal} {\bibinfo  {journal} {Proceedings of the IEEE}\ }\textbf {\bibinfo
  {volume} {93}},\ \bibinfo {pages} {216} (\bibinfo {year} {2005})}\BibitemShut
  {NoStop}%
\bibitem [{\citenamefont {Born}\ and\ \citenamefont
  {Wolf}(1980)}]{BornWolf1980}%
  \BibitemOpen
  \bibfield  {author} {\bibinfo {author} {\bibfnamefont {M.}~\bibnamefont
  {Born}}\ and\ \bibinfo {author} {\bibfnamefont {E.}~\bibnamefont {Wolf}},\
  }\href@noop {} {\emph {\bibinfo {title} {{Principles of Optics:
  Electromagnetic Theory of Propagation, Interference and Diffraction of
  Light}}}},\ \bibinfo {edition} {6th}\ ed.\ (\bibinfo  {publisher}
  {Pergamon},\ \bibinfo {year} {1980})\BibitemShut {NoStop}%
\bibitem [{\citenamefont {Gourlay}(2011)}]{gourlay2011}%
  \BibitemOpen
  \bibfield  {author} {\bibinfo {author} {\bibfnamefont {M.~R.}\ \bibnamefont
  {Gourlay}},\ }\href {\doibase 10.1007/978-90-481-2639-2_164} {\emph {\bibinfo
  {title} {Encyclopedia of Modern Coral Reefs: Structure, Form and Process}}},\
  edited by\ \bibinfo {editor} {\bibfnamefont {D.}~\bibnamefont {Hopley}}\
  (\bibinfo  {publisher} {Springer Netherlands},\ \bibinfo {year}
  {2011})\BibitemShut {NoStop}%
\bibitem [{\citenamefont {Zolotovskii}\ \emph {et~al.}(2018)\citenamefont
  {Zolotovskii}, \citenamefont {Dadoenkova}, \citenamefont {Moiseev},
  \citenamefont {Kadochkin}, \citenamefont {Svetukhin},\ and\ \citenamefont
  {Fotiadi}}]{zolotovskii2018}%
  \BibitemOpen
  \bibfield  {author} {\bibinfo {author} {\bibfnamefont {I.~O.}\ \bibnamefont
  {Zolotovskii}}, \bibinfo {author} {\bibfnamefont {Y.~S.}\ \bibnamefont
  {Dadoenkova}}, \bibinfo {author} {\bibfnamefont {S.~G.}\ \bibnamefont
  {Moiseev}}, \bibinfo {author} {\bibfnamefont {A.~S.}\ \bibnamefont
  {Kadochkin}}, \bibinfo {author} {\bibfnamefont {V.~V.}\ \bibnamefont
  {Svetukhin}}, \ and\ \bibinfo {author} {\bibfnamefont {A.~A.}\ \bibnamefont
  {Fotiadi}},\ }\href {\doibase 10.1103/PhysRevA.97.053828} {\bibfield
  {journal} {\bibinfo  {journal} {Phys. Rev. A}\ }\textbf {\bibinfo {volume}
  {97}},\ \bibinfo {pages} {053828} (\bibinfo {year} {2018})}\BibitemShut
  {NoStop}%
\bibitem [{\citenamefont {Morgado}\ and\ \citenamefont
  {Silveirinha}(2017)}]{morgado2017}%
  \BibitemOpen
  \bibfield  {author} {\bibinfo {author} {\bibfnamefont {T.~A.}\ \bibnamefont
  {Morgado}}\ and\ \bibinfo {author} {\bibfnamefont {M.~G.}\ \bibnamefont
  {Silveirinha}},\ }\href {\doibase 10.1103/PhysRevLett.119.133901} {\bibfield
  {journal} {\bibinfo  {journal} {Phys. Rev. Lett.}\ }\textbf {\bibinfo
  {volume} {119}},\ \bibinfo {pages} {133901} (\bibinfo {year}
  {2017})}\BibitemShut {NoStop}%
\bibitem [{\citenamefont {Castro}\ \emph {et~al.}(2010)\citenamefont {Castro},
  \citenamefont {Ochoa}, \citenamefont {Katsnelson}, \citenamefont {Gorbachev},
  \citenamefont {Elias}, \citenamefont {Novoselov}, \citenamefont {Geim},\ and\
  \citenamefont {Guinea}}]{castro2010}%
  \BibitemOpen
  \bibfield  {author} {\bibinfo {author} {\bibfnamefont {E.~V.}\ \bibnamefont
  {Castro}}, \bibinfo {author} {\bibfnamefont {H.}~\bibnamefont {Ochoa}},
  \bibinfo {author} {\bibfnamefont {M.~I.}\ \bibnamefont {Katsnelson}},
  \bibinfo {author} {\bibfnamefont {R.~V.}\ \bibnamefont {Gorbachev}}, \bibinfo
  {author} {\bibfnamefont {D.~C.}\ \bibnamefont {Elias}}, \bibinfo {author}
  {\bibfnamefont {K.~S.}\ \bibnamefont {Novoselov}}, \bibinfo {author}
  {\bibfnamefont {A.~K.}\ \bibnamefont {Geim}}, \ and\ \bibinfo {author}
  {\bibfnamefont {F.}~\bibnamefont {Guinea}},\ }\href {\doibase
  10.1103/PhysRevLett.105.266601} {\bibfield  {journal} {\bibinfo  {journal}
  {Physical Review Letters}\ }\textbf {\bibinfo {volume} {105}},\ \bibinfo
  {pages} {266601} (\bibinfo {year} {2010})}\BibitemShut {NoStop}%
\bibitem [{\citenamefont {Tomadin}\ \emph {et~al.}(2014)\citenamefont
  {Tomadin}, \citenamefont {Vignale},\ and\ \citenamefont
  {Polini}}]{tomadin2014}%
  \BibitemOpen
  \bibfield  {author} {\bibinfo {author} {\bibfnamefont {A.}~\bibnamefont
  {Tomadin}}, \bibinfo {author} {\bibfnamefont {G.}~\bibnamefont {Vignale}}, \
  and\ \bibinfo {author} {\bibfnamefont {M.}~\bibnamefont {Polini}},\ }\href
  {\doibase 10.1103/PhysRevLett.113.235901} {\bibfield  {journal} {\bibinfo
  {journal} {Phys. Rev. Lett.}\ }\textbf {\bibinfo {volume} {113}},\ \bibinfo
  {pages} {235901} (\bibinfo {year} {2014})}\BibitemShut {NoStop}%
\bibitem [{\citenamefont {Chen}(2012)}]{chen2012}%
  \BibitemOpen
  \bibfield  {author} {\bibinfo {author} {\bibfnamefont {F.~F.}\ \bibnamefont
  {Chen}},\ }\href
  {https://www.amazon.com/Introduction-Plasma-Physics-Francis-Chen/dp/1475704615?SubscriptionId=AKIAIOBINVZYXZQZ2U3A&tag=chimbori05-20&linkCode=xm2&camp=2025&creative=165953&creativeASIN=1475704615}
  {\emph {\bibinfo {title} {Introduction to Plasma Physics}}}\ (\bibinfo
  {publisher} {Springer},\ \bibinfo {year} {2012})\BibitemShut {NoStop}%
\end{thebibliography}%

\end{document}